\documentclass[reprint,superscriptaddress,amsmath,amssymb,aip,floatfix]{revtex4-2}
\usepackage{pdfpages,pgffor,graphicx,dcolumn,bm,xr,physics,chemformula,siunitx,color,soul,float,appendix, subfigure, booktabs,array, multirow}

\colorlet{blue}{blue!70!black} 
\colorlet{red}{red!60!black}

\def\bk{{\bf k}}
\def\br{{\bf r}}

\def\bq{{\bf q}}

\def\bR{{\bf R}}

\def\bT{{\bf T}}

\def\a{\alpha}

\def\k{\kappa}

\def\ve{\varepsilon}
\def\btau{\boldsymbol{\tau}}

\def\D{\partial}
\def\d{\delta}
\def\<{\langle}
\def\>{\rangle}

\def\hH{{\hat{H}}}

\makeatletter
\AtBeginDocument{\let\LS@rot\@undefined}
\makeatother

\begin{document}

\raggedbottom

\title{Comparison between first-principles supercell calculations of polarons and the \textit{ab initio} polaron equations\vspace{10pt}}

\author{Zhenbang Dai}
\affiliation{Oden Institute for Computational Engineering and Sciences, The University of Texas at Austin, Austin, Texas 78712, USA}
\affiliation{Department of Physics, The University of Texas at Austin, Austin, Texas 78712, USA}

\author{Donghwan Kim}
\affiliation{Oden Institute for Computational Engineering and Sciences, The University of Texas at Austin, Austin, Texas 78712, USA}
\affiliation{Department of Physics, The University of Texas at Austin, Austin, Texas 78712, USA}

\author{Jon Lafuente-Bartolome}
\affiliation{Department of Physics, University of the Basque 
         Country UPV/EHU, 48940 Leioa, Basque Country, Spain}

\author{Feliciano Giustino}%
\email{fgiustino@oden.utexas.edu}
\affiliation{Oden Institute for Computational Engineering and Sciences, The University of Texas at Austin, Austin, Texas 78712, USA}
\affiliation{Department of Physics, The University of Texas at Austin, Austin, Texas 78712, USA}
        
\date{\today}

\begin{abstract}
Polarons are composite quasiparticles formed by excess charges and the accompanying lattice distortions in solids, and play a critical role in transport, optical, and catalytic properties of semiconductors and insulators. The standard approach for calculating polarons from first principles relies on density functional theory and periodic supercells. An alternative approach consists of recasting the calculation of polaron wavefunction, lattice distortion, and energy as a coupled nonlinear eigenvalue problem, using the band structure, phonon dispersions, and the electron-phonon matrix elements as obtained from density functional perturbation theory. Here, we revisit the formal connection between these two approaches, with an emphasis on the handling of self-interaction correction, and we establish a compact formal link between them. We perform a quantitative comparison of these methods for the case of small polarons in the prototypical insulators \ch{TiO2}, MgO, and LiF. We find that the polaron wavefunctions and lattice distortions obtained from these methods are nearly indistinguishable in all cases, and the formation energies are in good (\ch{TiO2}) to fair (MgO) agreement. We show that the residual deviations can be ascribed to the neglect of higher-order electron-phonon couplings in the density functional perturbation theory approach. 
\end{abstract}
\maketitle

\section{Introduction}
An excess electron or hole in a semiconductor or insulator induces a local deformation of the surrounding lattice. This distortion can in turn cause the localization of the excess charge. The resulting quasiparticle is named polaron~\cite{Landau_1932a, Devreese_2020, Franchini_Diebold_2021}, and is important in  transport~\cite{Zhang_Venkatesan_2007, Celiberti_Franchini_2024, Zhang_Svistunov_2023}, catalytic and photocatalytic properties~\cite{Gong_Diebold_2006, Reticcioli_Franchini_2019, Setvin_Diebold_2014}, surface reconstruction~\cite{Reticcioli_Franchini_2017}, and optical properties~\cite{Emin_1993, Miyata_Zhu_2017, Mishchenko_Svistunov_2003} in a wide range of materials.

In order to gain insight into polaron behaviors in real materials, the standard method consists of performing density functional theory (DFT) calculations of supercells containing an excess charge (electron or hole). This method is convenient since it can be used with any electronic structure code, but faces two challenges: (i) the supercell needs to be large enough to fully encompass the polaron wave function and to minimize the interaction between its periodic images; this requirement can make calculations prohibitive owing to the cubic scaling of DFT~\cite{Giustino_2014}. (ii) The local and semilocal exchange-correlation functionals used in DFT give rise to spurious polaron self-interaction error, which tends to favor delocalization and hinders polaron formation~\cite{Mori_Yang_2008, Sio_Giustino_2019b, Falletta_Pasquarello_2022a}. 
The first challenge can be addressed by using finite-size correction schemes~\cite{Freysoldt_VandeWalle_2014, Falletta_Pasquarello_2020}; overcoming the second challenge requires the use of polaron self-interaction correction (pSIC) schemes. 

\begin{figure*}
    \centering
    \includegraphics[width=0.5\linewidth]{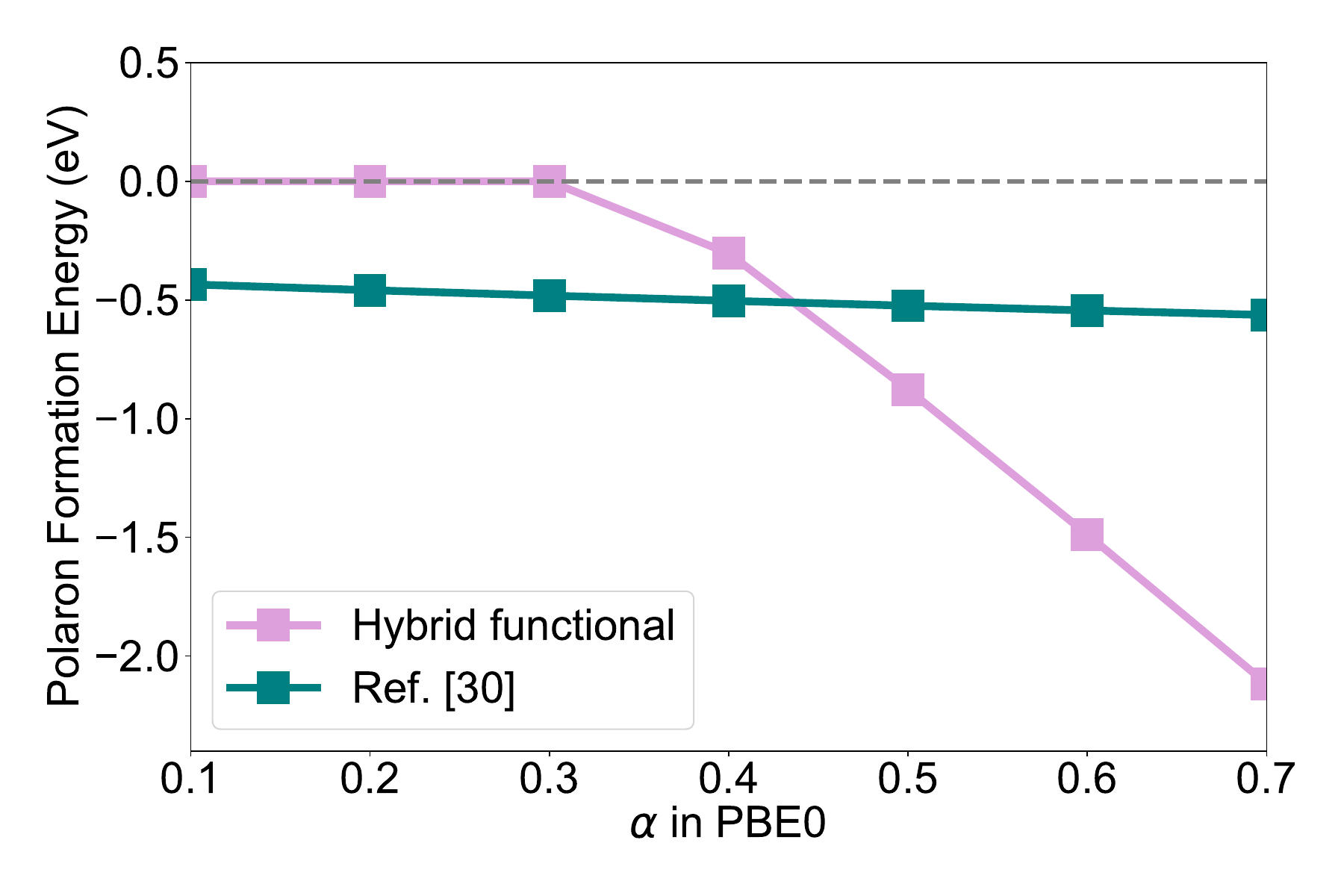}
    \caption{Supercell calculations of the formation energy of the small one-center hole polaron in \ch{LiF}. The pink symbols are from PBE0 hybrid functional calculations of a charged supercell, as a function of the fraction of exact exchange $\a$. The green symbols are from the pSIC method of Ref.~\citenum{Sadigh_Aberg_2015} with the PBE0 functional, and correspond to charge-neutral supercells; also in this case we perform calculations for varying $\a$. For both methods and for all datapoints, the structure of the hole polaron is the same and is fixed to that obtained from the \textit{ab initio} polaron equations~\cite{Sio_Giustino_2019a}, using the PBE functional and a 3$\times$3$\times$3 supercell.
    }
    \label{Fig:hybrid}
\end{figure*}

The use of DFT+$U$~\cite{Deskins_Dupuis_2007, Himmetoglu_Cococcioni_2014} or hybrid functionals~\cite{Elmaslmane_McKenna_2018, Kronik_Baer_2012, Heyd_Ernzerhof_2003, Perdew_Bruke_1996} leads to a partial cancellation of the self-interaction error, but the results tend to be sensitive to the choice of the Hubbard parameter or the fraction of exact exchange~\cite{Setvin_Diebold_2014, Kokott_Scheffler_2018}. For example, Fig.~\ref{Fig:hybrid} shows how different choices of the fraction of exact exchange in the PBE0 functional~\cite{Perdew_Bruke_1996} can give either localized polarons or no polarons at all. Furthermore, in the case of hybrid functionals, calculations for large supercells may be impractical~\cite{Lin_2016}. 

An alternative strategy for overcoming the self-interaction error while avoiding tunable parameters consists of designing new DFT functionals specifically for polarons that are self-interaction free by design. These approaches have successfully been demonstrated for small polarons~\cite{Lany_Zunger_2009, Avezac_Mauri_2005, Falletta_Pasquarello_2022a, Sio_Giustino_2019b, Sadigh_Aberg_2015}. Among these approaches, the pSIC scheme introduced by Sadigh \textit{et al.}\cite{Sadigh_Aberg_2015} is especially advantageous since it only requires standard DFT calculations in charge-neutral supercells, as discussed in Sec.~\ref{sec:sea} of this manuscript. As such, this method does not require the modification of the DFT energy functional or existing electronic structure codes.

In addition to DFT supercell calculations, it is now possible to study polarons starting from density functional perturbation theory (DFPT)~\cite{Sio_Giustino_2019a, Sio_Giustino_2019b,Lee_Bernardi_2021,Luo_Bernardi_2022,Luo_Bernardi_2025,Vasilchenko_Gonze_2022}. These approaches find their roots in studies of polarons based on effective Hamiltonians such as the ones in the classic Pekar, Fr\"ohlich, and Holstein models~\cite{Pekar_1946, Feynman_1955, Lee_Pines_1953, Holstein_1959a, Holstein_1959b, Lang_Firsov_1963}. In these approaches, the problem of finding polaron formation energy and wavefunction is formulated in terms of the electron bands, the phonon dispersions, and the electron-phonon coupling matrix elements; these methods are based on the approximations of harmonic lattice and linear electron-phonon couplings. Starting from these ingredients, several possibilities have been investigated: variational minimization of the energy~\cite{Sio_Giustino_2019a, Sio_Giustino_2019b, Vasilchenko_Gonze_2022}, canonical transformations~\cite{Lee_Bernardi_2021, Luo_Bernardi_2022}, Green's function methods~\cite{Lafuente_Giustino_2022a, Lafuente_Giustino_2022b}, and diagrammatic Monte Carlo~\cite{Luo_Bernardi_2025}. The main advantage of these approaches is that the calculations of DFT and DFPT quantities are performed in the crystal unit cell, and the DFT supercell is replaced by an equivalent grid of electron and phonon wavevectors in the Brillouin zone; to investigate larger supercell one only needs denser Brillouin zone grids. The connection between these methods and standard DFT calculations is provided by the \textit{ab initio} polaron equations introduced by \textcite{Sio_Giustino_2019a}; these equations led to the identification of large polarons in alkali halides~\cite{Sio_Giustino_2019b}, two-dimensional materials~\cite{Sio_Giustino_2023}, halide perovskites~\cite{Lafuente_Giustino_2024}, and transition metal oxides~\cite{Dai_Giustino_2024c}. Beyond the significant computational saving as compared to DFT supercell approaches, these equations offer a detailed view of which electrons, phonons, and electron-phonon couplings drive the formation of polarons; furthermore this method naturally connects to many-body effective Hamiltonian approaches to polarons~\cite{Lafuente_Giustino_2022a, Lafuente_Giustino_2022b}. On the other hand, this method does not contain anharmonic lattice dynamics and nonlinear electron-phonon couplings, which are captured in full DFT supercell calculations.

Despite the successes of DFT supercell calculations of polarons and the reciprocal-space polaron equations, several outstanding questions about these methods remain. At a practical level, there is a need for detailed benchmark studies comparing these approaches within the same computational settings and materials palette, especially for small polarons where the harmonic and linear approximations may not apply. At a conceptual level, there is a need for a more in-depth analysis of the formal relation between these methods. 

Here, we begin to fill these gaps by performing a comparison between the pSIC method of \textcite{Sio_Giustino_2019b}, the the supercell approach of Sadigh \textit{et al.}\cite{Sadigh_Aberg_2015}, and the \textit{ab initio} polaron equations of \textcite{Sio_Giustino_2019a}. In particular, we derive a compact expression for the polaron formation energy which unites these three methods under a single formalism, and we show how this unified approach also connects to the GW method~\cite{Hybertsen_Louie_1986}.
Furthermore, we perform calculations of small polarons in LiF, MgO, and anatase \ch{TiO2} using both methods by Sadigh \textit{et al.}~\cite{Sadigh_Aberg_2015} and reciprocal-space \textit{ab initio} polaron equations~\cite{Sio_Giustino_2019b}, and we show that the results for polaron wavefunction, atomic displacements, and energetics are in fair/good agreement; in particular, the formation energies differ by as little as 2\% in \ch{TiO2} and as much as 36\% in LiF; while the lattice distortion differ by as little as 17\% in anatase \ch{TiO2} and as much as 28\% in LiF. The differences are ascribed to the lack of nonlinear electron-phonon couplings in the reciprocal-space method, and highlight the importance of developing methods to compute second- and higher-order electron-phonon coupling matrix elements in DFPT. 

The manuscript is organized as follows. In Sec.~\ref{sec:dftenergy} we outline the notation used in this manuscript, in Sec.~\ref{sec:sio} we discuss the pSIC method of \textcite{Sio_Giustino_2019b}, reformulate it into a more compact expression, and in Sec.~\ref{sec:sea} we show how the approach introduced by Sadigh \textit{et al.}\cite{Sadigh_Aberg_2015} can immediately be derived from the results of Sec.~\ref{sec:sio} by making one simple approximation. We further discuss in this section the connection with the GW method. In Sec.~\ref{sec:connection} we discuss how the \textit{ab initio} polaron equations of \textcite{Sio_Giustino_2019b} can in turn be obtained from the method of Sadigh \textit{et al.}\cite{Sadigh_Aberg_2015} by introducing the approximations of harmonic lattice and linear electron-phonon couplings. We report detailed numerical benchmark tests for LiF, MgO, and \ch{TiO2} in Sec.~\ref{sec:calculations}. In Sec.~\ref{sec:conclusion} we offer our conclusions and indicate avenues for future work. We leave some technical aspects to the appendices. In particular, in App.~\ref{App:hybrid} we derive the method of Sadigh \textit{et al.}\cite{Sadigh_Aberg_2015} for hybrid functionals, and in App.~\ref{App:computation} we provide details of our computational setup.

\section{DFT total energy for polarons} \label{sec:dftenergy}

In this section we introduce the notation that will be used in the remainder of the manuscript. We consider the DFT total energy of a system in the presence of a polaron. For ease of notation we consider DFT with local or semilocal exchange and correlation functionals, such as for example the local density approximation~\cite{Ceperley_Alder_1980} or the generalized gradient approximation~\cite{Perdew_Ernzerhof_1996}. In App.~\ref{App:hybrid} we generalize this analysis to the case of hybrid functionals.

For definiteness we consider an electron polaron, but the following analysis holds unchanged for hole polarons. We assume that the system without polaron is gapped, contains $N$ electrons in the valence band manifold, and is spin-unpolarized; the system with polaron contains $N+1$ electrons, and has one unpaired electron spin. We use $\btau_0$ to indicate the entire set of ionic coordinates in the crystal without polaron, and $\btau$ to denote a distorted configuration; $\btau$ may refer to any configuration of the crystal, including one with a localized polaron. The electrons and ions are in a periodic Born-von-K\'arm\'an (BvK) supercell.

We use the following notation to denote Kohn-Sham (KS) wavefunctions in three scenarios: (i) For the $N$-electron system in the undistorted configuration $\btau_0$, we denote the KS states as $\psi^{N}_{v\bk \sigma}(\br; \btau_0)$. The index $v$ runs over the valence bands, which are fully occupied, $\bk$ is the electron wavevector, and $\sigma = \uparrow,\downarrow$ is the electron spin. These KS states are normalized in the BvK supercell, which consists of $N_\bk$ primitive unit cells. The total number of electron is $N = 2 N_{\rm v} N_\bk$. (ii) For the $N$-electron system in the distorted configuration $\btau\ne \btau_0$, we indicate KS states via $\psi^{N}_{v\sigma}(\br; \btau)$. In this case, we do not specify the wavevector label since the crystal periodicity is lifted by the distortion; the index $v$ runs from 1 to $N/2$, and the density is 
$n^{N}_{{\rm v},\sigma}(\br; \btau) = \sum_{v} |\psi^{N}_{v\sigma}(\br; \btau)|^2$. (iii) For the system with the excess electron, we denote KS states within the valence band manifold as $\psi^{N+1}_{v\sigma}(\br; \btau)$, and the wavefunction of the excess electron belonging to the conduction manifold as $\psi_{\rm p}(\br; \btau)$. 

The electron density in the distorted $(N\!+\!1)$-electron system can be decomposed into the up- and down-spin channels of the valence band, as well as the density of the extra electron:
 \begin{equation}
   n^{N+1}(\br; \btau)= n^{N+1}_{\rm v,\uparrow}(\br; \btau) + n^{N+1}_{\rm v,\downarrow}(\br; \btau)+
      n_{\rm p,\uparrow}(\br; \btau)~,
 \end{equation}
where $n^{N+1}_{{\rm v},\sigma}(\br; \btau) = \sum_{v} |\psi^{N+1}_{v\sigma}(\br; \btau)|^2$ and $n_{{\rm p},\uparrow}(\br; \btau) = |\psi_{\rm p}(\br; \btau)|^2$. The excess electron is taken to have spin-up without loss of generality. Using these definitions, the DFT total energy of the system can be written as~\cite{Giustino_2014}:
\begin{widetext}
\vspace{-10pt}
\begin{align}
\label{Eq:tot_energy_w_si}
    &E^{N+1}(\btau) =
    -\frac{\hbar^2}{2m} \sum_{v\sigma} \!\int \!\!d\br \abs{\nabla \psi^{N+1}_{v\sigma}(\br;\btau)}^2
    \hspace{-2  pt}-\hspace{-2pt}
    \frac{\hbar^2}{2m} \!\int \!\!d\br \abs{\nabla \psi_{\rm p}(\br;\btau)}^2
    \hspace{-2pt}+\!\int \!\!d\br\! \left[n^{N+1}_{\rm v}(\br;\btau) + n_{\rm p,\uparrow}(\br;\btau) \right]
    V_\mathrm{n}(\br;\btau)
    + E_\mathrm{n{\text -}n}(\btau)
    \nonumber \\
    &\!\!+\frac{1}{2}\frac{e^2}{4\pi\varepsilon_0}\!\sum_\bT\!\int \!\! d\br\, d\br' 
    \frac{\left[n_{\rm v}^{N+1}(\br;\btau) + n_{\rm p,\uparrow}(\br;\btau)\right] 
    \!\!\left[n_{\rm v}^{N+1}(\br';\btau) + n_{\rm p,\uparrow}(\br';\btau)\right]}{\abs{\br-\br'+\bT}}
    \!+\!E_\mathrm{xc}\!\!\left[n^{N+1}_{{\rm v},\uparrow}(\br;\btau) + n_{\rm p,\uparrow}(\br;\btau), n^{N+1}_{{\rm v},\downarrow}(\br;\btau)\!\right]\!,
\end{align}
\vspace{-10pt}
\end{widetext}
where $\hbar$, $m$, $e$, and $\varepsilon_0$ are the usual Planck constant, electron mass, electron charge, and vacuum permittivity, respectively; $V_{\rm n}$ is the electron-nuclear potential energy, which in pseudopotential codes is replaced by the sum of the ionic pseudopotentials; $E_\mathrm{n{\text -}n}$ denotes the nucleus-nucleus or ion-ion electrostatic interaction energy of the BvK supercell, $E_{\rm xc}$ is the exchange and correlation energy, and $n^{N+1}_{\rm v} = n^{N+1}_{{\rm v},\uparrow}+n^{N+1}_{{\rm v},\downarrow}$. The first two terms in this expression represent the kinetic energy, while the first term on the second third line is the Hartree energy and involves a summation over the BvK lattice vectors $\bT$. In Eq.~\eqref{Eq:tot_energy_w_si}, the integrals are taken over the BvK supercell, and $E^{N+1}$ refers to the energy of this supercell. To maintain charge neutrality, a uniform compensating positive background charge must be added to Eq.~\eqref{Eq:tot_energy_w_si}; since this term drops out in the following derivations~\cite{Sio_Giustino_2019b}, we omit it for brevity. 

\section{Polaron self-interaction correction by Sio \textit{et al.} (Ref. 16)}\label{sec:sio}

Using the notation introduced in Sec.~\ref{sec:dftenergy}, we now review the pSIC approach of \textcite{Sio_Giustino_2019b} and we recast it into a more compact expression that will be used in Secs.~\ref{sec:sea} and \ref{sec:connection} to derive the methods of Sadigh \textit{et al.}\cite{Sadigh_Aberg_2015} and \textcite{Sio_Giustino_2019a}.

Inspection of Eq.~\eqref{Eq:tot_energy_w_si} shows that there are two sources of polaron self-interaction: the first one is in the Hartree energy term, and is given by the Coulomb repulsion between the excess electron and itself:
\begin{align}
    \label{Eq:Hartree_SI}
    \frac{1}{2}\frac{e^2}{4\pi\varepsilon_0}\!\int \!d\br \,d\br' 
    \frac{n_{\rm p}(\br;\btau) n_{\rm p}(\br';\btau)}{\abs{\br-\br'}}~.
\end{align}
This term should not be present in exact DFT. The second source of self-interaction is in the exchange and correlation term, as it can be seen by performing a functional Taylor expansion of the energy $E_{\rm xc}$ in Eq.~\eqref{Eq:tot_energy_w_si}:
\begin{align}
    \label{Eq:xc_SI}
    &\hspace{-10pt}E_\mathrm{xc}[n^{N+1}_{{\rm v},\uparrow}(\br;\btau) + n_{\rm p}(\br;\btau), n^{N+1}_{{\rm v},\downarrow}(\br;\btau)]=
    \nonumber \\
    &\hspace{-5pt}
    E_\mathrm{xc}[n^{N+1}_{{\rm v},\uparrow}(\br;\btau), n^{N+1}_{{\rm v},\downarrow}(\br;\btau)]
    +
    \!\int \!\!d \br \,\frac{\d E_\mathrm{xc}}{\d n_{{\rm v},\uparrow}} n_{\rm p}(\br;\btau)
    \nonumber \\
    &\hspace{-5pt}+
    \frac{1}{2}\int d \br d \br' \frac{\d^2 E_\mathrm{xc}}{\d n^{N+1}_{{\rm v},\uparrow} \d n^{N+1}_{{\rm v},\uparrow}} n_{\rm p}(\br;\btau) n_{\rm p}(\br';\btau)~,
\end{align}
which is correct to order $(n_{\rm p})^2$. In this expression, the derivatives are evaluated at the valence-only spin densities of the $(N\!+\!1)$-electron system, $n^{N+1}_{{\rm v},\sigma}(\br;\btau)$. The term with the first-order derivative represents the exchange and correlation interaction between this valence density and the excess electron. The term with the second derivative is the spurious self-interaction of the excess electron, and should not be present in exact DFT. Higher-order terms in the expansion involve additional self-interaction effects, which we neglect in the following discussion.

To eliminate these self-interaction errors, \textcite{Sio_Giustino_2019b} introduced the following SIC scheme:
\begin{align}
    \label{Eq:Sio_pSIC}
    &\hspace{-6pt}\Delta E_{\rm pSIC} = -E_{\rm H}[n_{\rm p}] 
    - \frac{1}{2}\big\{E_{\rm xc}[n^{N+1}_{{\rm v},\uparrow}+n_{\rm p}, n^{N+1}_{{\rm v},\downarrow}]
    \nonumber \\
    &\hspace{-2pt}- 2 E_{\rm xc}[n^{N+1}_{{\rm v},\uparrow}, n^{N+1}_{{\rm v},\downarrow}]\big\}
    +E_{\rm xc}[n^{N+1}_{{\rm v},\uparrow}-n_{\rm p}, n^{N+1}_{{\rm v},\downarrow}]~,
\end{align}
which is a generalization of earlier work by d'Avezac \textit{et al.}\cite{Avezac_Mauri_2005}.
The first term in this energy correction removes the Hartree self-interaction error of Eq.~\eqref{Eq:Hartree_SI} exactly. In addition, it removes the interaction between the excess electron and its periodic images. The terms within curly braces represent the central finite-differences formula for the second derivative of the exchange-correlation energy; these terms cancel the exchange-correlation kernel term appearing in the last line of Eq.~\eqref{Eq:xc_SI}, up to second order in the density of the excess electron. The self-interaction corrected energy is obtained by summing Eq.~\eqref{Eq:tot_energy_w_si} and \eqref{Eq:Sio_pSIC}, 
\begin{equation}\label{Eq:toten_sio}
E_{\rm pSIC}^{N+1}(\btau) = E^{N+1}(\btau)+\Delta E_{\rm pSIC}(\btau)~.
\end{equation}
Here, we go one step further with respect to the method of Ref.~\citenum{Sio_Giustino_2019b}. We observe that, upon expanding $E_{\rm xc}$ to second order in $n_{\rm p}$, the energy $E_{\rm pSIC}^{N+1}$ can conveniently be recast in the highly compact form: 
\begin{align}
    \label{Eq:tot_energy_wo_si}
    &\hspace{-6pt}E_{\rm pSIC}^{N+1}(\btau) = E[n^{N+1}_{{\rm v},\uparrow}(\btau), n^{N+1}_{{\rm v},\downarrow}(\btau);\btau]  \nonumber \\
    &\hspace{-2pt}+\bra{\!\psi_{\rm p}(\btau)} 
    \!\hH_{\rm KS}[n^{N+1}_{{\rm v},\uparrow}(\btau), n^{N+1}_{{\rm v},\downarrow}(\btau);\btau]
    \!\ket{\psi_{\rm p}(\btau)\!},
\end{align}
having neglected terms of third order and higher in the density $n_{\rm p}$ of the excess electron. 
In this expression, $E[\cdots]$ is the same as in Eq.~\eqref{Eq:tot_energy_w_si}, except that here it is evaluated for the spin densities $n^{N+1}_{{\rm v},\sigma}(\btau)$, which \textit{do not} include the excess electron. Similarly, $\hH_{\rm KS}$ is the KS Hamiltonian, evaluated again for the spin densities $n^{N+1}_{{\rm v},\sigma}(\btau)$ without excess electron. Equation~\eqref{Eq:tot_energy_wo_si} provides a total energy which is free from polaron self-interaction, and does not suffer from the delocalization problem and the sensitivity to the Hubbard parameter or the fraction of exact exchange illustrated in Fig.~\ref{Fig:hybrid}.

As it stands, Eq.~\eqref{Eq:tot_energy_wo_si} is not advantageous for practical calculations, because its evaluation would require the knowledge of $n^{N+1}_{{\rm v},\sigma}(\btau)$ and $\psi_{\rm p}(\btau)$, which in turn necessitate the minimization of Eq.~\eqref{Eq:toten_sio}, as already reported in Ref.~\citenum{Sio_Giustino_2019b}. Nevertheless, this equation provides a very useful starting point for subsequent approximation, and in particular for deriving in the most compact form the method of Sadigh \textit{et al.}\cite{Sadigh_Aberg_2015} and the reciprocal-space polaron equations of \textcite{Sio_Giustino_2019a}. We perform these steps in Secs.~\ref{sec:sea} and \ref{sec:sio}, respectively.

\section{Polaron self-interaction correction by Sadigh et al. (Ref. 30)}\label{sec:sea}

The polaron self-interaction correction of Ref.~\citenum{Sadigh_Aberg_2015} can be obtained from Eq.~\eqref{Eq:tot_energy_wo_si} by making the following formal replacement:
\begin{equation}\label{Eq:sadigh-approx}
n^{N+1}_{{\rm v},\sigma}(\btau) \,\to\, n^{N}_{{\rm v},\sigma}(\btau)~,
\end{equation}
i.e., by taking the valence charge density of the $N$-electron system in the distorted configuration $\btau$ as a proxy for the exact valence density of the $(N\!+\!1)$-electron system in the presence of the excess electron. This replacement involves two key approximations:
\begin{itemize}
\item[(i)] It is assumed that the excess electron does not modify the valence charge density of the $N$-electron system; this approximation corresponds to neglecting the effect of the Hartree, exchange, and correlation potentials generated by the excess electron on the valence manifold. In Ref.~\citenum{Sio_Giustino_2019b}, this approximation was motivated by noting that the excess electron contributes negligibly to the charge density in the limit of large supercell [cf.~\ discussion after Eq.~(10) of Ref.~\citenum{Sio_Giustino_2019b}]. 
\item[(ii)] If the $N$-electron system is spin-unpolarized, Eq.~\eqref{Eq:sadigh-approx} implies that the valence electrons of the $(N\!+\!1)$-electron system remain spin-unpolarized; this approximation corresponds to neglecting the spin-polarization of the valence manifold induced by the excess electron.  
\end{itemize}
In principle, these two approximations could be tested by evaluating the change of the valence density of the $N$-electron system induced by $V_{\rm H}[n_{\rm p}]+V_{\rm xc}[n_{\rm p}]$ in first-order perturbation theory. What one would obtain is that the error scales with the inverse of the band gap since the density variation only involves occupied-to-empty virtual transitions~\cite{Baroni_Gianozzi_2001}. In practice, it is easier to quantify the error by comparing the energy obtained within this approximation with the pSIC functional (without approximation) of Eq.~\eqref{Eq:toten_sio}.

Using Eq.~\eqref{Eq:sadigh-approx} inside Eq.~\eqref{Eq:tot_energy_wo_si}, one finds the compact expression of Sadigh \textit{et al.}\cite{Sadigh_Aberg_2015}:
\begin{align}
    \label{Eq:plrn_etot_Sadigh_0}
    &\hspace{-5pt}E_{\rm pSIC}^{N+1}(\btau) = E[n^{N}_{{\rm v}}(\btau);\btau]  \nonumber \\
    &\hspace{10pt}+\bra{\!\psi_{\rm p}(\btau)} 
    \!\hH_{\rm KS}[n^{N}_{{\rm v}}(\btau);\btau]
    \!\ket{\psi_{\rm p}(\btau)\!}~.
\end{align}
Note that we did not specify the spin densities separately since in the following discussion we will take the $N$-electron system to be spin-unpolarized.

When the density $n^{N}_{{\rm v}}(\btau)$ in Eq.~\eqref{Eq:plrn_etot_Sadigh_0} is set to the ground state density of the $N$-electron system, the minimum of $E_{\rm pSIC}^{N+1}(\btau)$ corresponds to its variational minimum with respect to the wavefunction $\psi_{\rm p}(\btau)$. This wavefunction must be normalized and orthogonal to all valence states of the $N$-electron system, $\psi^{N}_{v\sigma}(\btau)$. Imposing these conditions leads to the KS equation
\begin{align}
    \label{Eq:plrn_eq_Sadigh}
    &\hH_{\rm KS}[n^{N}_{{\rm v}}(\btau);\btau]\!\ket{\psi_{\rm p}(\btau)\!} = 
    \ve_{\rm CBM}(\btau) \!\ket{\psi_{\rm p}(\btau)\!}~,
\end{align}
where $\ve_{\rm CBM}$ denotes the $(N\!+\!1)$-th eigenvalue, i.e, the conduction band minimum. By combining Eqs.~\eqref{Eq:plrn_etot_Sadigh_0}-\eqref{Eq:plrn_eq_Sadigh}, one finds:
\begin{align}
    \label{Eq:plrn_etot_Sadigh}
    E_{\rm pSIC}^{N+1}(\btau) = E[n^{N}_{{\rm v}}(\btau);\btau] +\ve_{\rm CBM}(\btau)~.
\end{align}
This expression constitutes the central result of the method of Ref.~\citenum{Sadigh_Aberg_2015}. It shows that, within the approximation of Eq.~\eqref{Eq:sadigh-approx}, the self-interaction-corrected total energy of the $(N\!+\!1)$-electron system can be expressed as the sum of the ground-state energy of the $N$-electron, charge-neutral system, and the eigenvalue of the conduction band bottom of the same charge-neutral system. In addition, Eq.~\eqref{Eq:plrn_eq_eform} shows that, in this approach, the polaron wavefunction corresponds to the wavefunction of the conduction band minimum of the charge-neutral $N$-electron system. In practice, Eqs.~\eqref{Eq:plrn_etot_Sadigh} and \eqref{Eq:plrn_eq_Sadigh} are advantageous because they \textit{do not} require calculations for charged supercells, and the self-interaction correction does not rely on the Hubbard parameter or the fraction of exact exchange. The counterpart of Eq.~\eqref{Eq:sadigh-approx} for hole polarons contains $-\ve_{\rm VBM}(\btau)$ instead of $+\ve_{\rm CBM}(\btau)$, where VBM indicates the valence band maximum.

In App.~\ref{App:hybrid}, we generalize these considerations to the case of hybrid functionals, and we show that Eqs.~\eqref{Eq:plrn_etot_Sadigh} and \eqref{Eq:plrn_eq_Sadigh} maintain their validity even for hybrids. In Fig.~\ref{Fig:hybrid} we demonstrate the use of this approach with the PBE0 functional, by comparing the formation energies of the hole polaron in LiF using (i) standard calculations with charged supercells, and (ii) using Eq.~\eqref{Eq:plrn_etot_Sadigh}. It is evident that the latter method is almost completely insensitive to the choice of the fraction of exact exchange.

Equation~\eqref{Eq:plrn_etot_Sadigh} also carries an interesting similarity with many-body GW calculations~\cite{Hedin_Lundqvist_1969}, as discussed in detail in Ref.~\citenum{Dai_Giustino_2025b}, Sec.~VII.A. In fact, the GW total energy of the $(N\!+\!1)$-electron system can be written as:
\begin{equation}\label{eq:gw}
E^{N+1, \rm GW}(\btau)=E^{N}(\btau) + E_{\rm CBM}^{\rm GW}(\btau)~,
\end{equation}
where $E_{\rm CBM}^{\rm GW}(\btau)$ is the electron addition energy, i.e., the GW quasiparticle energy. In many-body perturbation theory~\cite{Hybertsen_Louie_1986}, all quantities in this expression are evaluated at the electron density of the $N$-electron system, therefore Eq.~\eqref{eq:gw} reduces to Eq.~\eqref{Eq:plrn_etot_Sadigh} upon making the replacement:
\begin{equation}\label{eq:gw2}
E_{\rm CBM}^{\rm GW}(\btau)  \,\to\, \varepsilon_{\rm CBM}^{\rm GW}(\btau)~.
\end{equation}
The close relation between Eq.~\eqref{eq:gw} and Eq.~\eqref{Eq:plrn_etot_Sadigh} suggests that the self-interaction corrections provided by the methods of Sadigh \textit{et al.}\cite{Sadigh_Aberg_2015} and \textcite{Sio_Giustino_2019b} can be viewed as DFT approximations to the GW polaron energy.

In practical calculations, the equilibrium structure of the polaron needs to be obtained by minimizing the functional $E_{\rm pSIC}^{N+1}(\btau)$ in Eq.~\eqref{Eq:plrn_etot_Sadigh} with respect to the atomic coordinates $\btau$. To this end, one needs to compute the atomic forces via the derivatives of $E_{\rm pSIC}^{N+1}(\btau)$ evaluated from the Hellman-Feynman theorem. This operation amounts to using the standard DFT forces already implemented in every electronic structure code for the first term on the right-hand side of Eq.~\eqref{Eq:plrn_etot_Sadigh}, i.e., the total energy $E^{N}[n^{N}_{{\rm v}}(\btau);\btau]$ of the charge-neutral system. For the second term in the same equation, i.e., the eigenvalue $\ve_{\rm CBM}(\btau)$, one can use Janak's theorem~\cite{Janak_1978} and evaluate numerical derivatives of the total energy with respect to the total charge,
$\ve_{\rm CBM} = \lim_{\d\to 0} (E^{N+\d}-E^N)/\d$, and then obtain the forces as for the first term~\cite{Sadigh_Aberg_2015}. At the end of this procedure, the formation energy of the polaron is obtained as: 
\begin{equation}
    \label{Eq:plrn_eform_Sadigh}
    \Delta E_{\rm f} = E_{\rm pSIC}^{N+1}(\btau) - E_{\rm pSIC}^{N+1}(\btau_0)~.
\end{equation}
$\Delta E_{\rm f}$ quantifies the energy lowering that results from the electron localization into the polaronic state.

\section{Connection with the \textit{ab initio} polaron equations}\label{sec:connection}

In Ref.~\citenum{Sio_Giustino_2019b}, the \textit{ab initio} polaron equations were derived starting from the total energy expression of Eq.~\eqref{Eq:tot_energy_w_si} and the self-interaction correction in Eq.~\eqref{Eq:Sio_pSIC}. Here, we show that the compact expression by Sadigh \textit{et al.}\cite{Sadigh_Aberg_2015} provides an alternative, natural starting point for deriving the same set of equations.

To this end, we consider the pSIC total energy in Eq.~\eqref{Eq:plrn_etot_Sadigh_0}, and we perform a Taylor expansion of 
$E[n^{N}_{{\rm v}}(\btau);\btau]$ and $\hH_{\rm KS}[n^{N}_{{\rm v}}(\btau)]$ in the atomic displacements around the configuration $\btau_0$ of the undistorted crystal: $\btau = \btau_0+\Delta\btau$. For the total energy we perform a second-order expansion, which corresponds to the \textit{harmonic approximation}:
\begin{align}
    \label{Eq:harmonic}
    &\hspace{-10pt}E[n^{N}_{{\rm v}}(\btau);\btau] = 
    E[n^{N}_{{\rm v}}(\btau_0);\btau_0]\nonumber\\  
    &\hspace{-5pt}+\frac{1}{2}\!\! \sum_{\substack{\k p \a \\\k' p' \a'}} C^N_{\k p \a, \k' p' \a'}[n^{N}_{{\rm v}}(\btau_0);\btau_0] \Delta \tau_{\k p \a} \Delta \tau_{\k' p' \a'}~,
\end{align}
where $C^N_{\k p \a, \k' p' \a'}$ is the matrix of interatomic force constants of the $N$-electron system in its ground state, and the indices $\k$, $\a$, and $p$ denote the atom, Cartesian direction, and unit cell withing the BvK supercell, respectively~\cite{Giustino_2017}. For the KS Hamiltonian, we perform a first-rder expansion, which corresponds to the \textit{approximation of linear electron-phonon couplings}:
\begin{align}
    \label{Eq:linear_eph}
    \hspace{-8pt}\hH_{\rm KS}[n^{N}_{{\rm v}}(\btau);\btau] =
    \hH_{\rm KS}[n^{N}_{{\rm v}}(\btau_0);\btau_0] 
    + \sum_{\k p \a}\frac{\D \hH_{\rm KS}}{\D \tau_{\k p \a}} \Delta \tau_{\k p \a},
\end{align}
where the derivative is evaluated at the electronic density $n^{N}_{{\rm v}}(\btau_0)$ and the atomic configuration $\btau_0$.
The approximations of harmonic lattice and linear electron-phonon couplings are justified when polaronic distortions are small, and tend to become increasingly accurate for larger polarons. In Sec.~\ref{sec:calculations} we report quantitative tests of the range of validity of these approximations.

Upon replacing Eqs.~\eqref{Eq:harmonic} and \eqref{Eq:linear_eph} inside Eq.~\eqref{Eq:plrn_etot_Sadigh_0}, one obtains a variational optimization problem in the wavefunction $\psi_{\rm p}$ of the excess electron and in the atomic displacements $\Delta\btau$; the other quantities are fixed, pre-computed parameters. Minimizing the energy functional subject to the normalization condition for the electron wavefunction, one finds the \textit{ab initio} polaron equations in real space~\cite{Sio_Giustino_2019a}:
\begin{equation}
    \label{Eq:plrn_eqns}
    \left\{ \!\hH_{\rm KS}[n^{N}_{{\rm v}}(\btau_0);\btau_0] 
    \!+\! \sum_{\k p \a}\frac{\D \hH_{\rm KS}}{\D \tau_{\k p \a}} \Delta \tau_{\k p \a} \!\right\}\psi_{\rm p}
    = \ve_{\rm p} \psi_{\rm p}, 
\end{equation}
\begin{equation}
    \label{Eq:plrn_eqns2}
    \Delta \tau_{\k p \a} = 
    -\hspace{-6pt}\sum_{\k' p' \a'} C^{N,-1}_{\k p \a, \k' p' \a'} 
    \!\int\!\! d\br \frac{\D \hH_{\rm KS}}{\D \tau_{\k' p' \a'}}  |\psi_{\rm p}(\br)|^2.
\end{equation}
In practical calculations, these equations are more conveniently reformulated in reciprocal space; this is achieved by expanding the wavefunction $\psi_{\rm p}$ in a basis of KS state $\psi^N_{n\bk,\sigma}(\br;\tau_0)$ of the undistorted system, and the atomic displacements $\Delta\btau$ in a basis of vibrational eigenmodes~\cite{Sio_Giustino_2019b}. The reciprocal space versions of Eqs.~\eqref{Eq:plrn_eqns} and \eqref{Eq:plrn_eqns2} are provided in App.~\ref{App:computation} for completeness. 

By combining Eqs.~\eqref{Eq:harmonic}, \eqref{Eq:linear_eph}, \eqref{Eq:plrn_eqns}, and \eqref{Eq:plrn_eform_Sadigh}, the polaron formation energy in this approach takes the form:
\begin{equation}
    \label{Eq:plrn_eq_eform}
    \Delta E_{\rm f} 
    = \frac{1}{2} \sum_{\substack{\k p \a \\\k' p' \a'}} C^N_{\k p \a, \k' p' \a'} \Delta \tau_{\k p \a} \Delta \tau_{\k' p' \a'}  + \ve_{\rm p} - \ve_{\rm CBM}(\btau_0)~.
\end{equation}
This expression is valid for electron polarons; for hole polarons, we replace $\ve_{\rm p} - \ve_{\rm CBM}(\btau_0)$ by $- [\ve_{\rm p} - \ve_{\rm VBM}(\btau_0)]$.

In the same way as Eq.~\eqref{Eq:plrn_etot_Sadigh} can be understood as an approximation of the GW total energy in Eq.~\eqref{eq:gw}, one could derive the \textit{ab initio} polaron equations starting from the GW method. The result is that the DFPT electron-phonon matrix elements appearing in Eqs.~\eqref{Eq:plrneqn_rec} and \eqref{eqn:bmat_plrn} must be replaced by the corresponding matrix elements computed via GW perturbation theory (GWPT)~\cite{Li_Louie_2019}. These aspects are dicussed in Ref.~\citenum{Dai_Giustino_2025b}.

The derivation outlined in this section and in Sec.~\ref{sec:sea} can be summarized as follows:
\begin{itemize}
\item[(i)] The method of Sadigh \textit{et al.}\cite{Sadigh_Aberg_2015} can be conceptualized as obtained from the pSIC functional of \textcite{Sio_Giustino_2019b}, as given by Eqs.~\eqref{Eq:toten_sio}, by making the approximation that the valence electron density at fixed atomic configuration $\btau$ is not modified by the presence of the excess electron;
\item[(ii)] The \textit{ab initio} polaron equations of Ref.~\citenum{Sio_Giustino_2019a} can be conceptualized as obtained from the method of Sadigh \textit{et al.}\cite{Sadigh_Aberg_2015} by performing the additional approximations of harmonic lattice and linear electron-phonon couplings.
\end{itemize}
In Sec.~\ref{sec:calculations} we proceed to a quantitative comparison of these approaches.

\section{Numerical comparison between the method of Sadigh et al. (Ref. 28) and the \textit{ab initio} polaron equations}\label{sec:calculations}

In this section we report a quantitative comparison between the method of Ref.~\citenum{Sadigh_Aberg_2015} (``supercell method'' henceforth) and the \textit{ab initio} polarons equations of Ref.~\citenum{Sio_Giustino_2019a} (``polaron equations'' in the following).

While both methods are in principle valid for polarons of any size, supercell calculations of intermediate-size and large polarons are impractical; therefore, here we focus on small polarons, and test the polaron equations in a worst-case scenario. 

We consider small hole polarons in prototypical insulators, namely anatase \ch{TiO2}, \ch{MgO}, and \ch{LiF}, which have been studied extensively~\cite{Lafuente_Giustino_2022b, Dai_Giustino_2024c, Falletta_Pasquarello_2022b, Deskins_Dupuis_2007, Deskins_Dupuis_2009, Lee_Bernardi_2021, Luo_Bernardi_2022, Luo_Bernardi_2025, Kokott_Scheffler_2018, Elmaslmane_McKenna_2018, Reticcioli_Franchini_2017, Reticcioli_Franchini_2019}. 
In the following, we use $3\times3\times2$, $5\times5\times5$, and $3\times3\times3$ supercells for \ch{TiO2}, \ch{MgO}, and \ch{LiF}, respectively. 
We choose to not apply finite-size correction schemes~\cite{Freysoldt_VandeWalle_2014, Falletta_Pasquarello_2020} since our goal is to compare the energetics of the two methods rather than investigating the formation energy of polarons in the dilute limit. The computational setup used for these calculations is detailed in App.~\ref{App:computation}.

\begin{figure*}
    \centering
    \includegraphics[width=0.98\linewidth]{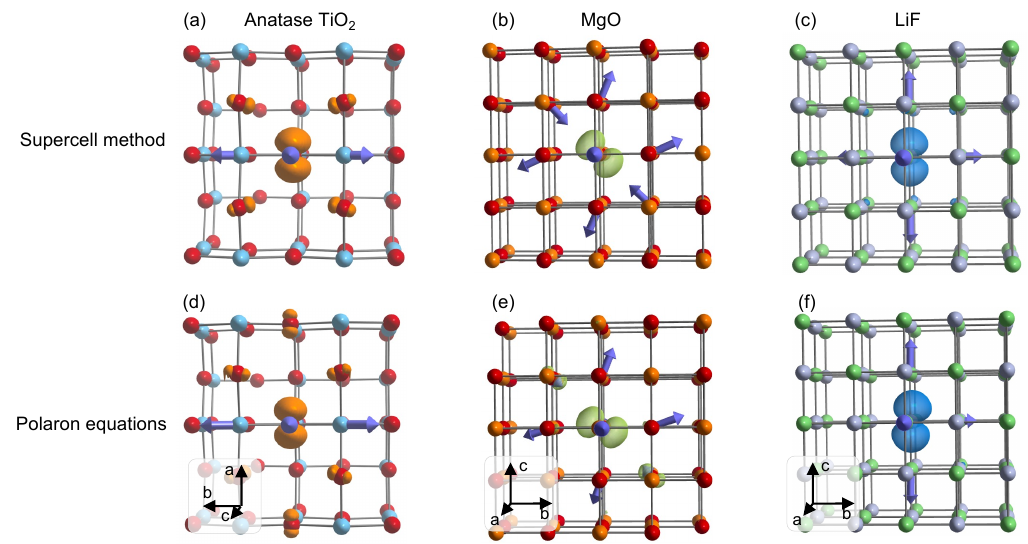}
    \caption{
    Comparison between the electron charge density and atomic displacements corresponding to the small hole polarons in anatase \ch{TiO2}, \ch{MgO}, and \ch{LiF}, as obtained from the supercell method of Ref.~\citenum{Sadigh_Aberg_2015} and from the \textit{ab initio} polaron equations of Ref.~\citenum{Sio_Giustino_2019a}.   
    (a) Ball-and-stick models of anatase \ch{TiO2} with Ti in light blue and O in red. The isosurface is the polaron charge density obtained from the supercell method, and the arrows indicate the corresponding atomic displacements with respect to the undistorted crystal. 
    For ease of visulization, the atomic displacements are magnified so that the maximum displacements computed from two methods coincide. 
    The differences of the actual atomic displacements can be found in Table~\ref{Tab:eform}.
    (d) Same as in (a), but obtained from the polaron equations.
    (b) and (e): Same as in (a) and (d), but for the hole polaron in MgO. Mg is shown in orange. 
    (c) and (f): Same as in (a) and (d), but for the one-center hole polaron in LiF. Li is in gray and F is in green.
    }
    \label{Fig:density}
\end{figure*}

Figure~\ref{Fig:density} shows ball-and-stick models of the polarons calculated with the supercell approach and with the polaron equations. The isosurfaces represent the charge density of the hole polaron, which in all cases corresponds to O-$2p$ or F-$2p$ orbitals; the arrows represent the atomic displacement from the undistorted crystal to the polaron ground state. We can see that the cations (Ti, Mg, Li) tend to move away from the polaron center, while the anions (O, F) are attracted toward it; this is consistent with the expectation from elementary electrostatic. Visual inspection of the panels in the top row of Fig.~\ref{Fig:density} (supercell method) and those in the bottom row (polaron equations) show that the two methods yield nearly indistinguishable solutions.

For a more quantitative comparison, we report in Tab.~\ref{Tab:eform} the polaron formation energies $\Delta E_{\rm f}$, the vertical excitation energies $\Delta \ve_{\rm p}=\ve_{\rm VBM}(\btau)-\ve_{\rm CBM}(\btau_0)$, and the change of nearest-neighbor bond lengths computed within each method. The vertical excitation energy represents the single-particle ionization energy of the polaron at fixed atomic configuration~\cite{Dai_Giustino_2025b}. This energy is typically significantly larger than the polaron formation energy; for reference, in simple polaron models such as the Landau-Pekar model~\cite{Pekar_1946}, one finds $\Delta \ve_{\rm p} = 3\Delta E_{\rm f}$.

In the case of \ch{TiO2}, the supercell method gives a formation energy of 298~meV, while the polaron equations yield 305~meV; therefore, in this case, the harmonic and linear approximations introduce an error of 7~meV or 2\% on the formation energy. 
The corresponding vertical excitation energies are also comparable, with the two methods yielding 929~meV and 1,215~meV, respectively. The variation of the nearest neighbor Ti-O bond length $d_{\rm TiO}$ and second-nearest neighbor O-O bond length $d_{\rm OO}$ are 0.118~\AA\ and $-0.044$~\AA\ for the supercell method, respectively; the corresponding values for the polaron equations are 0.138~\AA\ and $-0.029$~\AA. Here, the polaron equation yield more pronounced distortions for the nearest neighbor Ti-O bond by 17\%, in line with the higher formation energy.

In the case of MgO, Tab.~\ref{Tab:eform} shows that the polaron is rather shallow, with a formation energy of only 32~meV in the supercell method and 31~meV with the polaron equations, again showing the good agreement between the two approaches. 
In this case, the change in nearest neighbor and second-nearest neighbor Mg-O and O-O bond lengths $d_{\rm MgO}$ and $d_{\rm OO}$ are 0.082~\AA\ and $-0.028$~\AA\ for the supercell method, and $0.091$~\AA\ and $-0.022$~\AA for the polaron equations.
Here, we find that the polaron equations will also slightly overestimate the polaron distortions by 11\%.

For LiF, we find a larger discrepancy between the two methods, where the formation energy computed from the polaron equations overestimates that of the supercell method by 36\%, with the two methods yielding 887~meV and 652~meV, respectively. 
Similarly, the vertical excitation energies also have a larger error of 381 meV between the two methods.
Meanwhile, the larger error is manifested in the atomic displacements: The change in the nearest neighbor and second-nearest neighbor Li-F and F-F bond lengths $d_{\rm LiF}$ and $d_{\rm FF}$ are 0.206~\AA\ and $-0.043$~\AA\ for the supercell method, and $0.264$~\AA\ and $-0.026$~\AA\ for the polaron equations, in which the atomic distortion is overestimated by 28\%.

Taking together the results for \ch{TiO2}, MgO, and LiF, we can say that the polaron equations and supercell method provide polaron wavefunctions, lattice distortions, formation energies, and excitation energies that are generally in good agreement. Across these test systems, we find that the polaron equations tend to slightly overestimate the polaronic distortion and the vertical excitation energy.

\begin{figure*}
    \centering
    \includegraphics[width=0.9\linewidth]{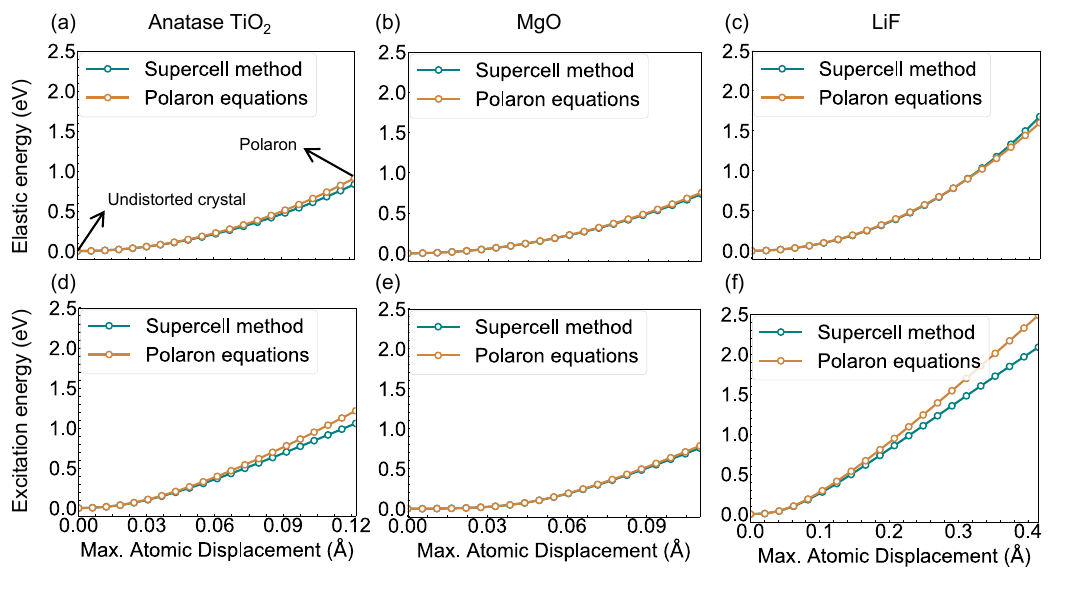}
    \caption{
    Decomposition of the potential energy surface of the small hole polarons in anatase \ch{TiO2} [panels (a) and (d)], \ch{MgO} [panels (b) and (e)], and \ch{LiF} [panels (c) and (f)] in terms of elastic energy (top row) and electronic excitation energy (bottom row), as defined in the main text. 
    The leftmost horizontal coordinate in each panel corresponds to the undistorted crystal, and the rightmost coordinate corresponds to the polaron structure, as shown by the arrows in (a). The horizontal axis indicates the maximum atomic displacement along the potential energy surface. Orange symbols are from the supercell method [Ref.~\citenum{Sadigh_Aberg_2015}], green symbols are from the polaron equations [Ref.~\citenum{Sio_Giustino_2019a}].
     }
    \label{Fig:higher_order}
\end{figure*}

In order to rationalize these quantitative differences between the supercell method and the polaron equations, in Fig.~\ref{Fig:higher_order} we compare the lattice and electronic contributions to the polaron potential energy surfaces. To this end, we consider a series of structures that are obtained by linearly interpolating between the undistorted crystal and the polaron structure as obtained from the polaron equations (the analysis remains unchanged if we consider the polaron structure from the supercell method as the final configuration); these structures are labeled by their maximum atomic displacement.
For each configuration, we show the change of the ground state energy, $E[n^{N}_{{\rm v}}(\btau);\btau]$ in Eq.~\eqref{Eq:plrn_eform_Sadigh}, and the change in electronic excitation energy, $\ve_{\rm CBM}(\btau)$ in Eq.~\eqref{Eq:plrn_eform_Sadigh}, referred to their values in the undistorted structure. The sum of these quantities gives the polaron formation energy according to Eq.~\eqref{Eq:plrn_eform_Sadigh}. In Fig.~\ref{Fig:higher_order}, we refer to these quantities as the ``elastic'' energy and the ``excitation'' energy, respectively. 

Visual inspection of Fig.~\ref{Fig:higher_order}(a)-(c) indicates that the two methods yield very similar elastic and excitation energies across the entire range of polaronic distortions. Furthermore, the elastic energies obtained from either method are in very good agreement, with a maximum deviation of 77~meV or 9\% in the case of \ch{TiO2}. These results indicate that the approximation of harmonic lattice is surprisingly effective, even for atomic displacements as large as 0.4~\AA\ as found for LiF. Clearly this finding is connected to the fact that all the test systems considered are known to be harmonic systems; this level of agreement may not hold for highly anharmonic crystals.

In addition, Fig.~\ref{Fig:higher_order}(d)-(f) show that, unlike the elastic energy, the excitation energy obtained from the polaron equations tend to overestimate that of the supercell method. This error is most pronounced in the case of LiF, where it amounts to 399~meV or 19\% at the polaron configuration. The impact of this error on the formation energy in the absolute scale is mitigated by the fact that the elastic energy and the excitation energy contribute with opposite signs, therefore there is partial error cancellation. Nevertheless, this discrepancy clearly indicates that the main source of deviation between the two methods is the lack of nonlinear electron-phonon couplings in the polaron equations.

Overall, the above comparison indicates that the supercell method and the polaron equations yield remarkably consistent results, and that a possible way forward to improve numerical agreement would be to introduce higher-order electron-phonon couplings in the polaron equations. We also note that the very small polarons analyzed in this section constitute a worst-case scenario for the harmonic and linear approximations at the basis of the polaron equations, and that an even closer agreement is expected for polarons involving two or more atomic orbitals.

\section{Conclusions}\label{sec:conclusion}

In summary, we have established a formal bridge between the supercell approach of Sadigh \textit{et al.}\cite{Sadigh_Aberg_2015} to polaron calculations and the \textit{ab initio} polaron equations of \textcite{Sio_Giustino_2019a}. Starting from the pSIC functional of Ref.~\citenum{Sio_Giustino_2019b} [Eq.~\eqref{Eq:Sio_pSIC}] and making a single, controlled approximation on the valence electron density, we recovered the formulation of Sadigh \textit{et al.}\cite{Sadigh_Aberg_2015} [Eq.~\eqref{Eq:plrn_etot_Sadigh}]. In turn, by making the approximations of harmonic lattice and linear electron-phonon coupling to the latter formalism, we derived the \textit{ab initio} polaron equations of \textcite{Sio_Giustino_2019a}. This chain of approximations clarifies the assumptions that underpin each approach, and provides a common language to compare them. The present formal analysis provides a complementary perspective on a related comparative study of these methods that was recently reported in Ref.~\citenum{Falletta_Pasquarello_2025}. 

We performed quantitative comparisons between these methods using the small hole polarons in anatase \ch{TiO2}, \ch{MgO}, and \ch{LiF} as test cases. The calculated polaron wavefunctions and atomic distortions are consistent in all cases; the formation energies agree to within a few percent in the best case (\ch{TiO2}) and deviate around 36\% in the worst case for LiF. 
By analyzing the potential energy surfaces obtained within each method, we established that the main source of deviation lies in the neglect of higher-order electron-phonon couplings in the polaron equations. 

The close agreement between the two methods in the case of extremely small polarons, which constitute a worst-case scenario for the \textit{ab initio} polaron equations, is remarkable, and suggests that the two methods will be in even better agreement for larger polarons involving more than a single atomic orbital.

The present study points to several interesting avenues for future work: (i) We expect that incorporating second- and higher-order electron-phonon matrix elements and, where needed, anharmonic effects will close the remaining gap between polaron equations and supercell calculations. (ii) Within supercell approaches, the effect of the frozen-valence approximation in the method of Ref.~\citenum{Sadigh_Aberg_2015} could be quantified by direct comparison to the fully variational pSIC functional of Ref.~\citenum{Sio_Giustino_2019b}. (iii) Within the \textit{ab initio} polaron equations, we expect that using GW band structures and GWPT electron-phonon matrix elements will further improve predictive accuracy. (iv) Fully predictive calculations of polarons will necessitate taking into account quantum nuclear effects and nonadiabaticity; progress in this direction has recently been made in the context of reciprocal-space approaches~\cite{Lafuente_Giustino_2022b, Luo_Bernardi_2022,Luo_Bernardi_2025}, and could be leveraged to enhance DFT supercell approaches.  

We hope that this work will serve as a starting point to make progress in each of these directions by bringing together the complementary strengths of each of these methods.

\begin{acknowledgments}
This work was supported by the Computational Materials Science program of the U.S. Department of Energy, Office of Science, Basic Energy Sciences under Award DE-SC0020129. Computational resources were provided by the National Energy Research Scientific Computing Center (a DOE Office of Science User Facility supported under Contract No.~DE-AC02-05CH11231), the Argonne Leadership Computing Facility (a DOE Office of Science User Facility supported under Contract DE-AC02-06CH11357), and the Texas Advanced Computing Center (TACC) at The University of Texas at Austin.
\end{acknowledgments}

\newpage

\begin{table*}[b]
\begin{tabular}{l@{\hspace{10pt}}r@{\hspace{10pt}}r@{\hspace{20pt}}r@{\hspace{10pt}}r@{\hspace{20pt}}r@{\hspace{10pt}}r}
\toprule
   & \multicolumn{2}{c}{\hspace{-20pt}\ch{TiO2}} & \multicolumn{2}{c}{\hspace{-20pt}MgO} & \multicolumn{2}{c}{\hspace{-20pt}LiF} \\[2pt] 
   & Supercell & Pol. Eqs. & Supercell & Pol. Eqs. & Supercell & Pol. Eqs. \\ 
\midrule
Formation energy (meV) & 298 & 305 & 32 & 31 & 652 & 887 \\
Excitation energy (meV) & 929 & 1215 & 755 & 786 & 2106 & 2487
 \\
Change of NN bond length (\AA) & 0.118 & 0.138 & 0.082 & 0.091 & 0.206 & 0.264 \\
Change of SNN bond length (\AA) & -0.044 & -0.029 & -0.028 & -0.022 & -0.043 & -0.026 \\
\bottomrule
\end{tabular}
\caption{Comparison between energetics and structure of the small hole polarons in anatase \ch{TiO2}, \ch{MgO}, and \ch{LiF}, as obtained from the supercell method (Ref.~\citenum{Sadigh_Aberg_2015}) and the polaron equations (Ref.~\citenum{Sio_Giustino_2019a}). We report the formation energy, the vertical excitation energy, and the change in the nearest neighbor (NN) and second-nearest neighbor (SNN) bond lengths. Bond lengths are averaged over equivalent atoms. These results \textit{do not} include finite-size supercell corrections. The computational setup is described in App.~\ref{App:computation}.}
\label{Tab:eform}
\end{table*}

\newpage

\newpage

\clearpage
\newpage

\appendix

\section{Generalization of the methods of Sadigh et al. (Ref. 28) and Sio et al. (Ref. 29) to hybrid functionals}
\label{App:hybrid}

Here we show that the methods of Secs.~\ref{sec:sea} and \ref{sec:connection} remain valid in the case of hybrid functional DFT calculations. For definitness we focus on the PBE0 functional~\cite{Perdew_Bruke_1996}, but the same reasoning extends to other hybrid functionals such as the HSE functionals~\cite{Heyd_Ernzerhof_2003}. As in Secs.~\ref{sec:sea} and \ref{sec:connection}, we consider a system with an excess electron in the spin-up channel.

When considering the PBE0 functional, Eq.~\eqref{Eq:tot_energy_w_si} must be modified as follows. 
First, we add the exact exchange term, scaled by the fraction $\a$:
\begin{align}
    \label{Eq:exchange}
    \a E_{\rm x} 
    =
    &-\frac{\alpha}{2}\frac{e^2}{4\pi\epsilon_0} 
    \sum_{vv'\sigma}
    \iint d\br d\br' 
    \frac{1}
    {|\br-\br'|}
    \nonumber \\
    &\hspace{30pt}\times 
    \psi_{v \sigma}^{N+1,*}(\br)\psi_{v' \sigma}^{N+1,*}(\br')\psi^{N+1}_{v'\sigma}(\br)\psi^{N+1}_{v \sigma}(\br')
    \nonumber \\
    &-
    \alpha\frac{e^2}{4\pi\epsilon_0} 
    \sum_{v } 
    \iint d\br d\br' 
    \frac{1}
    {|\br-\br'|} 
    \nonumber \\
    &\hspace{30pt}\times 
    \psi_{v \uparrow}^{N+1,*}(\br)\psi_{\rm p}^{*}(\br')\psi_{\rm p}(\br)\psi^{N+1}_{v \uparrow}(\br')
    \nonumber \\
    &-
    \frac{\a}{2}\frac{e^2}{4\pi\epsilon_0}  
    \iint d\br d\br' 
    \frac{n_{\rm p}(\br) n_{\rm p}(\br')}
    {|\br-\br'|}~.
\end{align}
In this expression, the parametric dependence of $\psi_{v\sigma}^{N+1}$ and $\psi_{\rm p}$ on the atomic configuration $\btau$ has been omitted for notational simplicity. Second, we scale the semilocal exchange in Eq.~\eqref{Eq:tot_energy_w_si} by $1\!-\!\a$. 

The last term in Eq.~\eqref{Eq:exchange} is of the same form as the Hartree energy in Eq.~\eqref{Eq:Hartree_SI}, but carries the opposite sign. Therefore this term tends to partially cancel the Hartree self-interaction of DFT calculations~\cite{Dai_Giustino_2025b}. When performing self-interaction correction, this term is removed from the total energy; similarly, the residual self-interaction associated with the semilocal exchange and correlation is removed, as discussed in relation to Eq.~\eqref{Eq:xc_SI}. After collecting all remaining terms, one finds the same result as in Eq.~\eqref{Eq:tot_energy_wo_si}, the only difference being that the total energy and the Hamiltonian now carry orbital dependence.

From this point, using the frozen-valence approximation one recovers Eq.~\eqref{Eq:plrn_eq_Sadigh}; by further making the approximations of harmonic lattice and linear electron-phonon couplings, one recovers Eq.~\eqref{Eq:plrn_eqns}. Therefore, both the method of Sadigh \textit{et al.}\cite{Sadigh_Aberg_2015} and that of \textcite{Sio_Giustino_2019a} can directly be used with hybrid functionals, without any change to the underlying methodology.

\section{Computational setup}
\label{App:computation}

\renewcommand\thefigure{S\arabic{figure}}
\setcounter{figure}{0}  

We perform DFT and DFPT calculations using the \textsc{Quantum Espresso} suite~\cite{Giannozzi_Baroni_2017}. We employ the generalized gradient approximation of Perdew, Burke, and Ernzerhof (PBE)~\cite{Perdew_Bruke_1996} to the DFT exchange and correlation functional, norm-conserving pseudopotentials~\cite{Hamann_2013, vanSetten_Rignanese_2018}, and a planewaves kinetic energy cutoff of 90~Ry. The initial structures of anatase \ch{TiO2}, \ch{MgO}, and \ch{LiF} are taken from the Materials Project database~\cite{Jain_Persson_2013}, and lattice vectors as well as atomic coordinates are subsequently optimized. 

To calculate polarons via the polaron equations, we employ primitive unit cells for \ch{LiF} and \ch{MgO}, and the conventional unit cell for anatase \ch{TiO2}. Electron-phonon couplings and polarons are calculated using the \textsc{EPW}~\cite{Lee_Giustino_2023} code, which calls the \textsc{Wannier90}~\cite{Mostofi_Marzari_2014} code in library mode to obtain maximally-localized Wannier functions. The reciprocal-space versions of Eqs.~\eqref{Eq:plrn_eqns} and \eqref{Eq:plrn_eqns2} are given by~\cite{Sio_Giustino_2019b}:
\begin{align}
    \label{Eq:plrneqn_rec}
    &\sum_{m'\bk'}
    \bigg[
    \varepsilon_{m\bk} \delta_{mm'} \delta_{\bk\bk'}
    -
    \frac{2}{N_\bk}\sum_{\nu} B_{\bk-\bk'\nu} g_{mm'\nu}(\bk', \bk-\bk')
    \bigg] 
    \nonumber \\ & \hspace{20pt}\times A_{m'\bk'}
    = \varepsilon A_{m\bk},  \\
    &B_{\bq \nu}  
    = 
    \frac{1}{N_\bk \hbar \omega_{\bq \nu}} 
    \sum_{\substack{mm' \bk'}}
    A^*_{m'\bk'} A_{m\bk'+\bq}
    g^*_{mm'\nu}( \bk', \bq),  \label{eqn:bmat_plrn}
\end{align}
where $n$, $\bk$, and $\varepsilon_{n\bk}$ are band index, crystal momentum, and eigenvalue of the KS state $\psi^{N}_{n\bk \sigma}(\br; \btau_0)$ of the $N$-electron system in the undistorted configuration $\btau_0$ (cf.\ Sec.~\ref{sec:dftenergy}), respectively. $\omega_{\bq \nu}$ denotes the frequency of the normal mode with branch index $\nu$ and wavevector $\bq$; these modes are obtained by diagonalizing the dynamical matrix constructed from $C^N_{\k p \a, \k' p' \a'}[n^{N}_{{\rm v}}(\btau_0);\btau_0]$. $g_{mm'\nu}(\bk, \bq)$ denotes the electron-phonon coupling matrix element between the KS states 
$\psi^{N}_{n\bk \sigma}(\br; \btau_0)$ and $\psi^{N}_{m\bk+\bq \sigma}(\br; \btau_0)$ via the change of the self-consistent potential associated with the phonon $\bq\nu$.
Once the solution vectors $B_{\bq\nu}$ are obtained, we determine the atomic displacements using:
\begin{equation}
    \label{eqn:disp_formula}
    \hspace{-7pt}\Delta \tau_{\k p \a}
    =
    -\frac{2}{N_\bk} \sum_{\bq \nu}
    B_{\bq \nu} \left(\frac{\hbar}{2M_{\kappa}\omega_{\bq \nu}}\right)^{\!\!1/2}
    \!\!\!\!e_{\kappa \alpha, \nu}(\bq) \,e^{i\bq\cdot\bR_p},  \!\!   
\end{equation}
where $M_{\kappa}$ is the mass of atom $\kappa$, $e_{\kappa \alpha, \nu}(\bq)$ is the phonon polarization vector; $\bR_p$ is the lattice vector of the $p$-th unit cell in the BvK supercell~\cite{Giustino_2017}. The polaron formation energy is given by~\textcite{Sio_Giustino_2019a}:
\begin{align}
    \Delta E_\mathrm{f} &= \frac{1}{N_\bk}\sum_{n\bk}\abs{A_{n\bk}}^2
    (\varepsilon_{n\bk} - \varepsilon^0_{\mathrm{CBM}})
    -
    \frac{1}{N_\bk}\sum_{\bq \nu} \abs{B_{\bq\nu}}^2 \hbar \omega_{\bq\nu},
\end{align}
where $\varepsilon^0_{\mathrm{CBM}}$ denotes the conduction band minimum in the undistorted structure without polaron.

In all cases, we use a uniform, 6$\times$6$\times$6 Brillouin zone grid for ground state calculations and for generating coarse-grid quantities needed by \textsc{EPW} for for Wannier-Fourier interpolation. Bands, phonons, and electron-phonon matrix elements are then interpolated onto 3$\times$3$\times$2, 5$\times$5$\times$5, and 3$\times$3$\times$3 $\bk$- and $\bq-$ grids for anatase \ch{TiO2}, \ch{MgO}, and \ch{LiF}, respectively. This choice corresponds to studying polarons in equivalent 3$\times$3$\times$2, 5$\times$5$\times$5, and 3$\times$3$\times$3 supercells, respectively.
24, 3, and 3 valence bands are included to construct and solve the polaron equations for the three materials.

Visualization of crystal structures, charge densities, and displacement patterns are performed using \textsc{Vesta}~\cite{Momma_Izumi_2011}.

\bibliography{literature}

\end{document}